\documentstyle[prb,preprint,aps]{revtex}
\begin{document}
\preprint{NSF-ITP-95-119}
\draft

\title{\Large{Time Periodic Behavior of Multiband Superlattices in Static
Electric Fields}}

\author{ Daniel W. Hone}
\address{Institute for Theoretical Physics and QUEST,
        University of California, Santa Barbara, CA 93106}
\author{ Xian-Geng Zhao$^*$}
\address{Institute for Theoretical Physics,
        University of California, Santa Barbara, CA 93106\\
Department of Physics, The University of Texas at Austin,
Austin, Texas 78712}

\date{\today}
\maketitle

\begin{abstract}
We use an analytic perturbation expansion for the two-band system of tight
binding electrons to discuss Bloch oscillations and Zener tunneling within
this model.  We make comparison with recent numerical results and predict
analytically the frequency of radiation expected from Zener
tunneling, including its disappearance, as a function of the system parameters.

\end{abstract}


\narrowtext
One of the earliest predictions of the theory of electrons in periodic
crystals is of Bloch oscillations --- time periodic motion at frequency
$\omega_B =  Fd/\hbar$ under application of a uniform constant field $F$ in a
crystal potential of lattice period $d$, associated with Bragg reflection at
Brillouin Zone boundaries.  But in ordinary atomic crystals, subjected to
accessible applied (electric) fields, scattering typically disrupts the
coherence of electronic motion in times very short compared to the Bloch
period $T_B = 2\pi/\omega_B$, preventing observation of the phenomenon.
However, the lattice constant $d$ of semiconductor superlattices, typically
two orders of magnitude larger, correspondingly reduces $T_B$ by the same
factor.  Esaki and Tsu~\cite{esaki} recognized 25 years ago, at the dawn of
this new technology, that this greatly improved the prospects of experimental
observation of Bloch oscillations and related phenomena.  Moreover, the
characteristic frequencies lie in the interesting far infrared region.  Not
surprisingly, then, there has been a great deal of subsequent activity, both
experimental~\cite{expt} and theoretical~\cite{theor}, in the study of high
quality semiconductor superlattices under the influence of static and time
periodic electric fields.  Transport and both linear and nonlinear optical
response have been of special interest.
In particular, Bloch oscillations have been shown\cite{coul} to survive the
addition of Coulomb interactions to the simple theory, and far infrared
radiation from such oscillations has now\cite{blochobs} been observed.

The one-dimensional periodicity of a quantum well superlattice gives as the
eigenstates for independent (noninteracting) electrons a set of ``minibands''
associated with motion in the growth direction, perpendicular to the wells.
For obvious reasons of simplicity, much of the theoretical study has been
limited to consideration of a single isolated miniband.  But, as emphasized
particularly in a recent Letter\cite{rjs} by Rotvig, Jauho and Smith (RJS),
there are essential interesting phenomena --- notably Zener tunneling ---
introduced by additional bands in the presence of a static electric field.
These authors reported a numerical study of the density matrix of a two-band
model, exhibiting a number of interesting types of periodic behavior of the
electrons.  It is the purpose of the present paper, (i) to make use of
analytic results in the form of an exact perturbation expansion to explore the
corresponding behavior throughout the space of the parameters characterizing
the two-band superlattice, and (ii) to make use of our analytically predicted
behavior to compare with selected numerical results in order to test the
convergence of the perturbation series so as to define its range of validity
for application to other problems and properties.  The structure of the series
suggests (correctly, we will show) that the convergence is much more rapid
than imposed by the obvious limits that can be set analytically.

The two-band model is, of course, the simplest extension beyond the single
band picture, and it does introduce those features characteristic of interband
communication.  But is it a reasonable approximation to any physical situation
to focus on a single pair of minibands while neglecting the higher bands which
inevitably also exist?  As we have pointed out before,\cite{honeholt} there is
at least one realistic situation where this should capture the dominant
physical behavior.  A superlattice can be grown dimerized, with the unit cell
consisting, {\it e.g.}, of a pair of quantum wells, separated from the adjacent
pair by a larger (wider) barrier than that which separates the two wells of
the basis pair.  Then the lowest level of an individual well is split within
the ``molecular'' pair.  These two levels form a pair of bands in the
superlattice which are well separated energetically from those arising from
higher levels in the well, and it is then reasonable to neglect all except
that lowest pair of bands.

We will use the standard two-band tight-binding Hamiltonian in a uniform
static electric field $E$:

\begin{eqnarray}
{\cal H} = &&\sum_n\bigg[(\Delta_a + neEd)a_n^\dagger a_n + (\Delta_b
+ neEd)b_n^\dagger b_n \nonumber\\
&&- (W_a/4)(a^\dagger_{n+1} a_n + h.c.)
+ (W_b/4)(b^\dagger_{n+1} b_n + h.c.)\nonumber\\
&& + eER(a^\dagger_n b_n
+ b^\dagger_n a_n)\bigg].
\label{HAM}
\end{eqnarray}
Here the subscripts label the lattice sites and the lower and upper minibands
are designated by symbols $a$ and $b$, respectively.  The first two terms
describe the site energies of the Wannier states in the presence of the
electric field, and $W_{a,b}$ are the widths of the isolated ($ E=0 $)
minibands induced by nearest neighbor hopping: $\epsilon^{a,b}(k) =
\Delta_{a,b} \mp (W_{a,b}/2) \cos (kd) $.  The last term is the on-site
electric dipole coupling between minibands; $eR$ is the corresponding dipole
moment.  This Hamiltonian, first used~\cite{fukuyama} by Fukuyama, Bari and
Fogedby in connection with this type of system, does neglect Coulomb
interactions and electric dipole elements between Wannier states on different
sites, but it contains the essential physics for the problem.  Note that the
hopping parameters $W_{a,b}$ are written here with opposite signs, so that
with both parameters positive the band structure at $E=0$ is of the standard
nearly free electron character, with direct band gaps at the zone boundary.
But the calculation to follow is valid for arbitrary signs of the parameters.

The application of a static electric field localizes all electronic energy
eigenstates, with a localization length which decreases with increasing
electric field strength $E$.  The energy spectrum associated with the
Hamiltonian~(\ref{HAM}) above becomes two interleaved Stark ladders:

\begin{equation}
\epsilon_{1,2}^m = \Omega_{1,2} + m\omega_B,
\label{eps12}
\end{equation}
where $\omega_B = eEd$ is the Bloch frequency defined above (we have taken
units with Planck's constant equal to unity), $m$ is an arbitrary integer, and
the Stark ladder offsets $\Omega_{1,2}$ depend on the field amplitude $E$.
The result so far can be established in several ways~\cite{fukuyama,zhao}.
A particularly
satisfying proof based on the symmetries evident in a suitably chosen gauge is
given in the Appendix.  But one of us has established,\cite{zhao} in addition,
an analytic solution which includes an explicit expression for the offset
between the two ladders $\Omega\equiv \Omega_2 - \Omega_1$.  As noted by RJS,
the Zener tunneling behavior undergoes dramatic changes when $\Omega$
approaches zero.   We first analyze that interplay between ordinary Bloch
oscillations and Zener tunneling with the use of those analytic results, to
understand how the behavior is affected by the parameters of the superlattice.
One of the most attractive features of the semiconductor systems is, after
all, that those parameters are to a considerable extent under the control
of the fabricator.

Formally, the analytic results are in the form of a pertubation series in the
ratio $R/d$ of the interband dipole coupling to the Bloch frequency.  It is
therefore natural to compare the exact energy eigenvalues to those of the
decoupled bands (for the Hamiltonian (\ref{HAM}) with $R=0$).  These are of
the same form as (\ref{eps12}):

\begin{equation}
\epsilon_0^{a,b}(n) = \Delta_{a,b} + n\omega_B,
\label{eps0}
\end{equation}
but the offsets $\Delta_{a,b}$ are now the bare energy parameters of the
Hamiltonian, independent of the applied field $E$.
These energies are plotted in Fig. 1, as in RJS, in the form
$(\epsilon_0^p - \Delta_a)/\omega_B$, which equals $n$ (for $p=a$) or
$(\Delta_a -\Delta_b)/\omega_B + n$ (for $p=b$), as a function of
$(\Delta_a -\Delta_b)/\omega_B$, giving a set of horizontal lines at the
integers and a
set of parallel lines of unit slope for bands $a$ and $b$, respectively.
Away from the crossing points of this unperturbed spectrum,
$\Delta_b-\Delta_a=n\omega_B$, where the two Stark ladders are degenerate, the
exact full energies are expected (and found) to show only small corrections
due to finite interband coupling $R$ for reasonable values of $R/d < 1$.
Similarly, the eigenstates are largely those of individual bands $a$ and $b$.
Then significant Zener tunneling is to be found only near the crossing points,
and we will concentrate on those special values of the field,
\begin{equation}
\omega_B \equiv eEd = (\Delta_b-\Delta_a) /\ell,
\label{res}
\end{equation}
where $\ell$ is an integer.  At these fields the eigenstates will be mixtures
of comparable weights of states in the two bands.  If the system is prepared
initially in one of the bands, Zener tunneling will exhibit oscillations of
the weight between the two bands with a dominant frequency of the Stark ladder
offset $\Omega$.  We will show, moreover, that there are special circumstances
where this offset vanishes and the tunneling disappears.

The results of Ref.\ \onlinecite{zhao} can be expressed for the
resonance condition of Eq.(\ref{res}) in the form
\begin{equation}
\Omega = {\omega_B\over \pi}\cos^{-1}\sum_{m=0}^\infty (R/d)^{2m}U^{(2m)}
[\ell,(W_b+W_a)/\omega_B],
\label{pert}
\end{equation}
where we have explicitly indicated the dependence of the terms $U^{(2m)}$ of
the perturbation series on the parameters of the system.  The contribution
$U^{(2m)}$ (other than the first: $U^{(0)}=1$) is a sum of $2^{2m}$ terms,
each of which is an ordered integral of the form
\begin{equation}
\int_0^{2\pi}\,dk_1\int_0^{k_1}\,dk_2\cdots\int_0^{k_{2m-1}}dk_{2m}
F_1(k_1)F_2(k_2)\cdots F_{2m}(k_{2m}),
\label{Uint}
\end{equation}
where each function $F_n(k)$ is of one of the two forms:
\begin{eqnarray}
C(k) = &&\cos ( \alpha\sin k - 2\ell k )\nonumber\\
S(k) = &&\sin ( \alpha\sin k - 2\ell k ).
\label{cs}
\end{eqnarray}
Here $\alpha = (W_b+W_a)/(2\omega_B)$, and the integer $\ell$ is again the
resonance parameter of Eq.~(\ref{res}).  The formal derivation of the
perturbation series shows that $|U^{(2m)}| < (2\pi)^{2m}/(2m)!$, so the
series is
convergent for all values of the ratio $R/d$.  On the other hand, the number
of terms and their algebraic complexity increase very rapidly with index
$(2m)$, so the result is of practical use only when it converges so rapidly
that one or two terms suffice.  In that regard the simple limit just invoked,
which puts an upper bound on the term of order $n$ of $(2\pi R/d)^n/n!$, is
not very helpful for typical values of $2\pi R/d$, which are of order unity.
But the highly oscillatory character of the integrands suggests that the
series converges, in fact, much more rapidly than this.  Our results below
verify that this is indeed the case.  We will keep only the first nontrivial
correction in $R/d$, namely the $m=1$ term, for which the integrals can
readily be expressed in closed form:
\begin{equation}
U^{(2)} = 2\left[\pi J_{\ell}(\alpha)\right]^2,
\label{u2}
\end{equation}
where $J_{\ell}(\alpha)$ is the Bessel function of order $\ell$.  Already we
see the substantial reductions from the oscillatory integrands from the crude
limit above, though we have been unable to establish rigorous lower analytic
limits than this.  Near $x=1$, $\cos^{-1}x$ has a square root singularity;
for $(R/d)^2U^{(2)}\ll 1$ we have as the offset between closest steps on the
two ladders,
\begin{eqnarray}
\Omega\approx &&\frac{\omega_B R}{\pi d}\sqrt{2U^{(2)}}\nonumber\\
= &&2\frac{(\Delta_b-\Delta_a)R}{\ell d}J_{\ell}(\alpha)
\label{omega}
\end{eqnarray}
We note, in particular, that this drops off rapidly with increasing $\ell$
(with decreasing field $E$ for fixed $(\Delta_b-\Delta_a)$, as defined in
(\ref{res})), or as we move to the right in the level crossings shown in
Fig. 1. This can be observed in the numerical results of Fig. 1 of RJS.  The
parameters used there are: $\Delta_b - \Delta_a = 20\rm\,meV$,
$R/d = -16/9\pi^2=-0.18$, and $\alpha=0.43\ell$, giving for the first three
gaps ($\ell = 1,2,3$) the values $\Omega = 1.51, 0.306, 0.030\rm\,meV$.  For
more direct comparison with the figure we reduce this by the Bloch frequency
in each case: $\Omega/\omega_B = 0.076, 0.031, 0.015$, which appears to be in
good agreement with those numerical results.

How accurate, in general, is the theory cut off at this lowest nontrivial
order?
To answer that, as well as to understand more easily the nature of the Zener
tunneling, it is useful to rederive the result (\ref{omega}) without the full
machinery of the exact solution, but rather using standard degenerate
perturbation theory.  For the independent band problem ($R=0$) the
Wannier-Stark eigenstates for band $a$ are given explicitly as
\begin{equation}
\phi_n^a = \sum_mJ_{n-m}\left(W_a\over 2\omega_B\right)a_m^{\dagger}|0\rangle,
\label{phia}
\end{equation}
with energy $\epsilon_0^a$ given by (\ref{eps0}).  The wave function for band
$b$ is, of course, of precisely the same form.  The resonance condition
(\ref{res}) then gives as degenerate the states $\phi_n^a$ and
$\phi_{n+\ell}^b$ for each integer $n$.  The degeneracy is lifted by the
interband dipole matrix elements, which are given explicitly between
degenerate unperturbed states as
\begin{eqnarray}
&&\langle \phi_n^a| {\cal H}_{int} |\phi_{n+\ell}^b\rangle \nonumber\\
&& = {R\over d}\omega_B\sum_mJ_{n-m}\biggl({W_a\over 2\omega_B}\biggr)
J_{n+\ell -m}\biggl({W_b\over 2\omega_B}\biggr) = {R\over d}\omega_B
J_{\ell}(\alpha).
\label{dipme}
\end{eqnarray}
Then the degenerate states are split by twice this amount in lowest order, in
agreement with (\ref{omega}), and the corresponding eigenstates are the
symmetric and antisymmetric combinations,
\begin{equation}
\psi_n^{s,a} = (\phi_n^a \pm \phi_{n+\ell}^b)/\sqrt{2}.
\label{sym}
\end{equation}
Now the nature of the Zener tunneling is transparent.  Let us prepare the
system in the (lower) band $a$ before turning on the field at time $t=0$.
Then the subsequent time development is given by
\begin{equation}
\psi(t) = \sum_n A_n\Bigl(\psi_n^s + \psi_n^ae^{-i\Omega t}\Bigr)
e^{-i(n\omega_B + \Omega_1)t},
\label{zener}
\end{equation}
where the symbols, other than the arbitrary constant coefficients $A_n$ which
determine the initial state, have all been defined above.  Clearly, the state
becomes completely made of contributions from band $b$ after time
$t = \pi/\Omega$ and returns to band $a$ in twice that time --- {\it i.e.},
with a full cycle of frequency $\Omega$, as seen in the numerical results of
RJS.  In particular, at the zeroes of the Bessel function,
$J_{\ell}(\alpha) = 0$, the splitting of the ladders $\Omega$ also vanishes,
and Zener tunneling also disappears.  Although this is tunneling between bands
rather than between the spatially separated sides of a standard double well,
the physical manifestations are not dissimilar.  Corresponding degenerate
states from the two bands are shifted by $\ell$ lattice sites, and there will
therefore be an ac current, or oscillating dipole associated with the Zener
tunneling, and corresponding radiation at the frequency $\Omega$.  Therefore,
the prediction of the theory to this order is that such radiation will be
observed at a frequency $\Omega \propto J_\ell(\alpha)$, except for the
special points where the Bessel function vanishes, at which the radiation
will disappear, as well.  These special points for  various resonance ratios
$\ell$ (electric field given by (\ref{res})) are given by
\begin{equation}
\alpha = {\ell(W_a+W_b)\over 2(\Delta_b-\Delta_a)} = x_\ell,
\label{zeros}
\end{equation}
where $x_\ell$ is a zero of $J_\ell(x)$.

The full theory predicts similar phenomena, but then the splitting of the
Stark ladders $\Omega$ needs to be calculated to all orders.  If the
corrections are significant, then the predictions of (\ref{omega}) and
(\ref{zeros}) will be at least quantitatively incorrect.

\acknowledgments

We thank Q. Niu, S.J. Allen, B.J. Keay, and G.H. D\"ohler
for stimulating discussions. This research was supported in part by the
National Science Foundation under Grant No. PHY94-07194. Additional support
was provided by NIST, the Robert A. Welch Foundation, the NNSF of China,
and the Grant LWTZ-1298 of Chinese Academy of Sciences.

{\appendix
\section{Proof of Interleaved Stark Ladder Spectrum}

The translational symmetry of the Hamiltonian (\ref{HAM}) is broken by the
scalar potential terms proportional to the intraband part of the position
operator,
\begin{equation}
x^0 \equiv d\sum_n n(a^{\dagger}_na_n + b^{\dagger}_nb_n).
\end{equation}
It is convenient to restore that translational symmetry by a suitable gauge
transformation (different from the standard vector potential gauge, because
only the intraband part of the position operator is involved):
\begin{equation}
\psi' = e^{ieEx^0t}\psi,
\end{equation}
which obeys the time-dependent Schr\"odinger equation $(i\partial_t -
{\cal H}') \psi' = 0$, with
\begin{eqnarray}
{\cal H}' = &&\sum_k {\cal H}'_k\nonumber\\
{\cal H}'_k = &&\Bigl[\Delta_a - (W_a/2)\cos(kd-\omega_Bt)\Bigr]
a^\dagger(k)a(k) \nonumber\\
+ &&\Bigl[\Delta_b + (W_b/2)\cos(kd-\omega_Bt)\Bigr] b^\dagger(k)b(k)
\nonumber\\
+ &&eER\Big[ a^\dagger(k)b(k) + b^\dagger(k)a(k)\Big].
\label{hprime}
\end{eqnarray}
Here $a(k)$, the spatial Fourier transform of $a_n$,
\begin{equation}
a(k) = \frac{1}{N}\sum_n e^{inkd}a_n,
\end{equation}
destroys a plane wave state of wave vector $k$ in band $a$, and the hermitian
conjugate operator $a^\dagger(k)$ creates that state. The operators $b(k)$ and
$b^\dagger(k)$ are defined analogously for the other band.  The dynamical
equation thus separates into a set of independent two-level problems, one for
each wave vector $k$, described by ${\cal H}'_k$.  Moreover,
${\cal H}'_k(t+T_B) = {\cal H}'_k(t)$, a discrete time translational symmetry
which, by the Floquet theorem, implies solutions of the form
\begin{equation}
\psi'_k(t) = u_k(t)\exp(-i\epsilon_kt),
\end{equation}
with $u_k(t) = u_k(t+T_B)$ periodic in time, and the ``quasienergy''
$\epsilon_k$ defined only modulo $\omega_B$.  Further, the form (\ref{hprime})
of ${\cal H}'_k$ makes it clear that $\epsilon_k$ and $u_k$ are, in fact,
independent of $k$: the parameter $k$ merely shifts the origin of time,
$t' = t - kd/\omega_B$ in each term.  Therefore, modulo $\omega_B$ there are
only two quasienergies, which we designate $\Omega_{1,2}$.  Finally, we can
see that this is also the set of {\it energies} of the system.  Transforming
back to the original time independent (scalar) gauge, we have
\begin{eqnarray}
\psi_k(t) = &&e^{-ieEx^0t}\Bigl[u^{(a)}(t)a^\dagger(k) + u^{(b)}(t)b^\dagger(k)
\Bigr]|0\rangle e^{-i\Omega_{1,2}t}\nonumber\\
= &&(1/N)\sum_n e^{in(kd-\omega_Bt)}\Bigl[u^{(a)}(t)a_n^\dagger + u^{(b)}(t)
b_n^\dagger\Bigr] |0\rangle e^{-i\Omega_{1,2}t},
\label{psik}
\end{eqnarray}
as the basis set of solutions to the dynamical Schr\"odinger equation in that
gauge ($u^{a,b}(t)$ are the parts of the Bloch periodic function $u(t)$
belonging to each of the two bands, in an obvious notation).  But these must
be linear combinations $\sum\phi_j\exp(-iE_jt)$ of energy eigenvalue solutions,
and we see from (\ref{psik}) that those energies are $E_j = \Omega_{1,2} +
n\omega_B$, for arbitary integers $n$ --- two interleaved Stark ladders of rung
spacing $\omega_B$, offset from each other by $\Omega = \Omega_2 - \Omega_1$.
}

\end{document}